\documentstyle[12pt]{article}
\newcommand{\be}{\begin{eqnarray}}

\newcommand{\ee}{\end{eqnarray}}

\title{
	\begin{flushright}
	{\normalsize
        TPI-MINN-97-12\\
        NUC-MINN-97-3-T\\
        UMN-TH-1541-96\\
        May 1997 \\}
	\end{flushright}
\bf     Small x Physics and Why It's Interesting       
       } 
\author{Larry McLerran\\ 
       {\small\it Theoretical Physics Institute, University of Minnesota, 
        Minneapolis, MN 55455  } \\          
       }

\date{}

\begin{document}
\maketitle

Abstract:  I discuss small x physics and its implications at very 
high energies.  At very high energy the density
of partons becomes so large that much of the physics can be described
using weak coupling methods in QCD.  This may allow for a solution to problems
such as the asymptotic nature of deep inelastic scattering at fixed
$Q^2$, unitarization, and multiparticle production.

\section{Some Old Problems We Would Like to Solve}

Small x physics is the study of the wee parton structure of
hadronic wavefunctions.  If we could understand this aspect of
hadrons, there are a variety of problems which we might understand:

\subsection{Total Cross Sections}

How do cross sections behave as the center of mass energy
$E \rightarrow \infty$? There are a variety of arguments 
based on unitarity which
suggest that the total $pp$ cross section at high energies behaves as
$\sigma \sim ln^2(E)$.  Is the coefficient for this $ln^2(E)$
behaviour universal and independent of what type of hadrons scatter?
In deep inelastic scattering at fixed $Q^2$ how does the cross section behave
at high energy?  Is this related in any simple way to hadronic cross sections?

\subsection{Cross Sections and Structure Functions at Small x}

Before addressing this issue, we should define $x$.  Let us work in 
a reference frame where a hadron has a very high longitudinal
momentum $P_h$.  Let the longitudinal momentum of a constituent be
$p$.  In this frame
\be
	x = p/P_h
\ee
It is of course possible to generalize this variable to a Lorentz invariant
variable, but working in the large momentum frame is conceptually useful.
We will do this generalization later.

Small x is of course the region where $x << 1$.  HERA measures the small
x part of parton structure functions.

We can look at this relationship for a component 
of the hadron wavefunction which
has small longitudinal momentum, almost at rest.  This hadron sees
the valence part of the hadron wavefunction as Lorentz contracted
to a scale size of order $x$.

In deep inelastic scattering, one measures the cross section for a photon of
virtuality $Q^2$ to scatter from a hadron as a function of energy.  Increasing
the energy corresponds to making $x$ smaller.  This cross section
is proportional to the structure functions. 

It is useful to introduce a rapidity variable
\be
	y = y_{hadron} - ln(1/x)
\ee
The variable $y_{hadron}$ can be chosen to be the smallest value
of x which is measured in a given experiment.  It is arbitrary in
this context.  If you like, it is the energy of the hadron 
in the large momentum frame.

The small x problem is the following:  
If we plot a parton distribution function
\be
	{{dN} \over {dy}} = x {{dN} \over {dx}}
\ee
as a function of y, then at small x, $dN/dy$ grows, perhaps as fast as
\be
	dN/dy \sim exp^{\kappa(y_{hadron}-y)}
\ee

The problem with this growth is it would seem 
that adding more constituents to
a hadron would increase the cross section at fixed $Q^2$.  This follows
since the cross section is proportional to the structure functions.  The
question we must ask is: How can the rapid small x growth seen 
at Hera be consistent
with the slow growth of hadronic cross sections expected from general
arguments based on unitarity? 

\subsection{Particle Production}

How are parton distribution functions related to particles which
are produced in hadron hadron interactions?  We know the answer to this
question for high transverse momentum jets which are produced in
hadron-hadron collisions.  What about the great majority of particles
which are not produced at huge values of transverse momenta?
Can we predict $dN/dy$ for produced hadrons as $y_{proj} \rightarrow
\infty$? 

What about fluctuations and correlations?  For example, does measuring
the two particle rapidity correlation function $dN/dy_1dy_2$ tell us anything
simple about the small x hadron wavefunction?  Is there a correlation
between transverse momenta and the rapidity density which one
can measure with $dN/dydp_T$?  

\subsection{Asymptotic Nature of the Quark Sea}

What is the intrinsic nature of the quark sea?  At small x,
what is the ratio of light quarks to glue?  
What is the ratio of heavy quarks to 
light?  What is the intrinsic transverse momentum of constituents of the sea?

\subsection{Initial Conditions for Heavy Ion Collisions}

Typically in heavy ion collisions, we describe the matter after some
formation time $t_{formation}$.  This is supposedly the time at which
particles are produced.  At this time, the initial conditions
for cascade and hydrodynamic simulations are formulated by some recipe.
Can one determine these initial conditions from first principles?

Before this time there is of course the wavefunction for the hadron.
If one understands this wavefunction, then surely one can determine the initial
conditions.  

At early times the problem is complicated by two types of coherence.
The first is quantum mechanical.  This is what forbids us to use
cascade simulations up to  $t = 0$ since in a cascade one 
simultaneously specifies
the momentum and coordinates of a participant in the cascade.  The second type
of coherence is charge coherence.  During the early stage of the collision,
the density of the hadron constituents 
of the nuclear wavefunction is very large,
$\sim \gamma$ where $\gamma$ is the nuclear Lorentz gamma factor.  However,
due to classical charge coherence, 
this high density cannot produce much effect. 
Because there is both positive and negative color charge,
disturbances at large wavelength cannot be generated.  This would not be the
case in a cascade where charge coherence is ignored (all scattering
is proportional to matrix elements squared).

\subsection{Universality}

Is there universal behaviour of cross sections at high energy?  
Are all measurable functions only of the local density of partons?
If we let $R$ be the radius of the hadron under consideration,
are all observables only functions of the local rapidity density per
unit area
\be
	\Lambda^2 = {1 \over {\pi R^2}} {{dN} \over {dy}}
\ee

\subsection{UNIVERSALITY}

Is the theory which describes the small x distribution functions a theory
at a critical point?  Are all observables determined by the universal
behaviour of this theory?  In the limit of $x \rightarrow 0$ are all
observables therefore determined by the symmetries of the theory which
describes these observables and {\bf UNIVERSALITY}?

\subsection{Claim}

{\bf MOST (MAYBE ALL) OF THE ABOVE MAY BE POSSIBLE TO
UNDERSTAND AS A RESULT OF THE HERA MEASUREMENTS OF SMALL X DISTRIBUTION
FUNCTIONS}

We will argue below that the results of the 
HERA experiments imply that the partons become very dense
at small x. This requires that the theory
of small x distribution functions has a small
strong coupling constant, $\alpha_S << 1$.  If this is the case, then
we should be able to compute properties of hadrons at small x from first 
principles in QCD.

\section{A Theoretical Sidebar}

Before proceeding further, we must develop more refined variables
for describing the physics at small x.  These will be light cone coordinates
and momenta.  If we let the energy be $P^0$ and longitudinal momentum
be $P^z$ then we define
\begin{eqnarray}
	P^\pm & = & {1 \over {\sqrt{2}}} (P^0 \pm P^z) \nonumber \\
        X^\pm & = & {1 \over {\sqrt{2}}} (t \pm z)
\end{eqnarray}
The dot product is
\be
	P \cdot X = p_T \cdot x_T - P^+ X^- - P^- X^+
\ee

We take the variable $P^-$ to be an energy variable.  The variable $X^+$
will therefore be the time variable conjugate to $P^-$.  States of the theory
will be eigenstates of $P^-$.

The longitudinal momenta will be taken to be $P^+$.  The longitudinal
spatial variable conjugate to $P^+$ will be $X^-$.

The uncertainty principle requires that
\be
	P^+ X^- \ge 1
\ee

The x variable may now be defined in a Lorentz invariant way using
light cone coordinates.  We take ratio of $p^+$ of a constituent to that
of the hadron $P^+_{hadron}$
\be
	x = p^+/P^+_{hadron}
\ee
It is easy to check that this is Lorentz boost invariant.

Two useful variables are the momentum space rapidity $y_{mom}$
and the space-time rapidity $y_{st}$.  We will construct these variables
from longitudinal momenta and coordinates as
\be
	y_{mom} = y_{hadron} - ln(1/x)
\ee
and
\be
	y_{st} = y_{hadron} - ln(x^- P^+_{hadron})
\ee

On account of the uncertainty principle, for the quantum mechanical state 
which describes small x physics, we expect that $x^- \sim 1/p^+$.
Therefore at small x where $y_{mom} >> 1$, we can take
\be
	y_{mom} \sim y_{st}
\ee 
that is the momentum space and space time rapidity must be equal up
to a unit or so of rapidity.  We will therefore use momentum space and 
space-time rapidity interchangeably and define
\be
	y = y_{st}
\ee

\section{The Only Scale in the Problem}

In hadron-hadron collisions, rapidity correlations are measured.
These correlations are measured by
\be
	{dN} \over {dy_1dy_2}
\ee
(The length of the correlation may grow slightly
with energy.)  It would be a miracle if there were correlations
in the wavefunction which did not appear in the final state distribution
of particles.  We will therefore assume that the wavefunction is more or 
less locally defined in rapidity.

The only variable therefore which is local which can describe this part of 
the hadron wavefunction is\cite{mv}
\be
	\Lambda^2 = {1 \over {\pi R^2}} {{dN} \over {dy}}
\ee
Here $R$ is the hadron radius.  Since the gluon density grows at small x,
we have
\be
	lim_{x \rightarrow 0}~ \Lambda^2 >> \Lambda_{QCD}^2
\ee
Therefore at small x, the strong coupling constant 
becomes small $\alpha_S (\Lambda) << 1$.

We arrive at one consequence of this trivially.  There should be universality
of physical quantities in terms of the parton density per unit area.

Because QCD is almost a scale invariant theory, the intrinsic
parton transverse momenta must grow at small x
\be
	p_T^2 \sim \Lambda^2
\ee

Since the intrinsic scale of $p_T$ grows, the sea must become flavor
symmetric.  Whenever $\Lambda >> m_{quark}$ the quark is essentially
massless and should contribute to the wavefunction as any light quark
would contribute.

\section{Hadrons vs Nuclei}

What is the fundamental difference between hadrons and nuclei at
small x?  Since everything depends only upon the parton density 
per unit area, there must be no fundamental difference.

There is a huge practical difference:  The rapidity distribution
is proportional to
\be
	dN/dy \sim A x^{-\delta}
\ee
where $\delta \sim .2 - .5$.  Therefore
\be
	\Lambda^2 \sim A^{1/3} x^{-\delta}
\ee
The gluon density at fixed x is much greater in a nucleus than
in a hadron.  We see that reducing x by 2 - 5 orders of magnitude
corresponds to using lead instead of a proton!

If the universality of parton distribution functions at small x is 
established, then using a nucleus to study small x is more
efficient than a proton.  This is because to get the same physics,
one must go to much smaller x in a proton, and therefore the energy
of an accelerator must be much higher.

\section{Transverse Momentum Broadening and Unitarity}

We can now understand how the constraint of unitarity is satisfied
in deep inelastic scattering.\cite{glr}  The deep inelastic scattering
cross section at fixed $Q^2$ is proportional to the parton density.
This is the integral over all partons with intrinsic transverse momenta
less than that $Q$,
\be
	xG(x,Q^2) = \int^{Q^2}_0 d^2p_T {{dN} \over {dy d^2p_T}}
\ee

The quantity which is computed from QCD is the density of
partons per unit area per unit $d^2p_T$, 
\be
	{1 \over {\pi R^2}} {{dN} \over {dy d^2p_T}}
\ee

At large $p_T$, the distribution of gluons is given by bremstrahlung
and should go as $1/p_T^2$.  There must be a coefficient
of $\Lambda^2$ in order to have the correct dimensions    
We can compute this coefficient in
the McLerran-Venugopalan model.\cite{mv} In this model, the sea quark density
follows the gluon density.  

At smaller values of $p_T$, the distribution softens,
to something which is at most logarithmically varying in
$p_T$.  On dimensional grounds, the $\Lambda $ dependence must also be weak.
The value of $p_T$ where $\Lambda^2 /p_T^2$ softens is $p_T \sim 
\Lambda$. 

We can now understand the x dependence of $xG(x,Q^2)$.  Suppose that
$Q^2 >> \Lambda^2$.  Then the integral above for $xG(x,Q^2)$
gets most of its contribution for $p_T >> \Lambda$.  $xG$ will
be proportional to $\Lambda^2$ and this will increase rapidly
as $x$ decreases.  This happens until $Q^2 \le \Lambda^2$.
At this point the integral is dominated by the small
$p_T$ region (but still much larger than $\Lambda_{QCD}$).
It has a very slowly varying dependence on $\Lambda$.
The fixed $Q^2$ cross section must therefore be slowly
rising.

The physics of this is easy to understand.  The small x enhancement 
corresponds to adding more higher $p_T$ components to the hadron wavefunction.
These components have smaller cross sections $\sim 1/p_T^2$.  We
are adding more constituents, but they are smaller and smaller
so that there is no conflict with unitarity.

\section{How Does a Parton See a Nucleus?}

The gluon field which is responsible for generating the gluon density
is produced by fluctuations in the color charge.  
These arise from sources with rapidity much greater than that
at which we are measuring the field.  Since the variation in the gluon
field is generated by interactions, the typical length scale for
variation in rapidity of the gluon density must be of
order $1/\alpha_S$.  This must also be the typical length scale over which
charges affect the field.  Since rapidity intervals 
of order $1/\alpha_S$ translate
into momentum ratios of order $e^{1/\alpha_S}$, the typical sources which
generate the gluon field must be Lorentz contracted down to infinitesmal
width.  Moreover since the sources move very close to the
speed of light, they should be effectively recoiless.

The typical transverse momentum is $p_T \sim \Lambda$.  This is
$p_T >> 1/R$.  therefore the concept of a hadron size is well defined.
Variations of the transverse shape of the distribution will
be very small compared to the typical parton wavelength.  We can
effectively treat the nucleus as of infinite transverse extent.

The source which generates the gluon field is therefore sitting on an 
infinitesmally thin sheet of infinite transverse extent 
propagating at the speed of light. 

Let us probe the distribution of color charge on the
surface.  On transverse resolution scales larger than a fermi,
the sheet is neutral.  On scales 
$\Delta x << 1 Fm$, one resolves sources coming
from individual nucleons.  These come from sources which are far separated 
in rapidity, and to a first approximation at least should be uncorrelated.
That is, the source is random and the fluctuations are controlled by a
Gaussian weight.

On transverse scales $\Delta x >> 1/\sqrt{\rho}$, where $\rho $ is the average
color charge squared from all sources at rapidity greater than the rapidity
at which we compute the field, 
\be
	(\Delta x)^2 \rho = Q^2 >> 1
\ee
This means that the color charge is in a high dimensional representation
of the color group and can be treated classically.

The theory which describes these fields is given by the path
integral\cite{mv}
\be
	\int [dA] [d \rho] exp\left(iS[A] + i A^-\rho
- {1 \over 2} \int dy d^2x_T \rho^2(y,x_T)/\mu^2(y) \right)
\ee
where here $\rho $ is the charge per unit area per unit rapidity.
The parameter $\mu^2(y)$ is the average charge squared per unit rapidity.
The total charge squared at rapidities larger than that of interest is
\be
	\chi = \int_y^{y_{proj}} dy^\prime \mu^2 (y^\prime)
\ee

This theory enables one to compute the intrinsic $p_T$ distribution.
It has the features described above in the section about unitarity in
deep inelastic scattering.  The parameter $\Lambda^2 \sim \chi$

The question to ask is what determines $\mu^2(y)$?\cite{mk}  
To understand this, we 
must recognize that the above theory is an effective theory valid only
for rapidities close to those for which we measure the field.  If
we compute quantum corrections to this theory, we get big corrections
of order $\alpha_S Y$ where $Y$ is a cutoff on the maximal allowed 
rapidity.

To compute the effective theory, we must integrate out the high rapidity modes.
This can be done in weak coupling if we sequentially integrate out
higher momentum modes to finally arrive at a low energy Lagrangean.
This procedure is the Wilson-Kadanoff renormalization group.
The Lagrangean above and $\mu^2(y)$ are determined by this procedure.
In fact higher order terms in $\rho$ might be generated as well
(although these are not important at $p_T \sim \Lambda$). 
All of these coefficients are determined by the renormalization group.
Perhaps the entire dependence of the structure functions on rapidity may
be determined by universality and the symmetries of this theory.

This renormalization group procedure at intermediate and large
$p_T \ge \Lambda$ is a linear equation.  It reduces in various limits to the
BFKL and DGLAP equations for the evolution of structure functions.\cite{eqn}
At small $p_T$, the evolution equations become non-linear and 
presumably saturate, that is, the distribution functions cease evolving.

\section{What Does the Gluon Field Look Like?}

The equations of motion for the gluon field are the non-abelian
Yang-Mills equations in the presence of a source localized in $x^-$.
The solution is easy to construct.  Suppose the field is a pure two
dimensional gauge transform of vacuum field for $x^- > 0$ 
and another gauge transform for $x^- < 0$
\be
	 A^i(x^-,x_T) = \theta(x^-) \alpha_1^i (x_T)
+ \theta (-x^-) \alpha_2^i (x_T)
\ee
where
\be
	\alpha^i_j(x_T) = i U_j (x_T) \nabla^i U^\dagger_j (x_T)
\ee
We take $A^\pm = 0$.

At $x^- \ne 0$, the solution has zero field strength.  The field
strength is concentrated at $x^- = 0$.
The discontinuity conditions and a boundary condition at say $x^- \rightarrow
-\infty$ determine the $U_j$.

The only big field strength is $F^{i+}$.  This field strength gives
$E \perp B$ with both perpendicular to the longitudinal direction.
These are the precise non-abelian analogs of the Lienard-Wiechert potentials
and Weizsacker-Williams fields of electrodynamics for a fast moving
source of electromagnetic charge.

\section{Hadron-Hadron Scattering}

From our knowledge of the non-abelian Lienard-Wiechert potentials, we can 
now construct solutions for the two hadron scattering problem.
\cite{mkw}-\cite{gyu}  Initially we 
have two infinitesmal sheets of charge approaching one another at the
speed of light.  On either side of the hadrons and in the region between
them we have three separate fields which are two dimensional
gauge transforms of vacuum.
At t = 0 there is a singularity of the equations of motion.  This form
of the solution no longer solves the equations of motions after the 
collision.

The solution which solves the equations of motion after $t=0$, and the
boundary conditions at $t=0$ is easy to construct.  If we
ignore a weak dependence on rapidity, the solution is
\be
	A^\pm & = & x^\pm \beta^\pm (\tau, x_T) \nonumber \\
        A^i & = & = \beta^i (\tau, x_T)
\ee
Here $\tau = \sqrt{t^2-x^2}$.  This is the solution in the forward 
light cone.

This solution has the property that distributions of particles are
Lorentz invariant.  It inevitably 
leads to hydrodynamics equations of the form
discussed by Bjorken.

At early times, the equations are non-linear.  The initial fields
are processed by these non-linearities.  At large $\tau >> 1/\Lambda$, 
the equations 
linearize.  This is because the form of these fields describes an
expanding matter distribution.  When the equations become
linear, one has plane waves of gluons.  These form the initial conditions
for a cascade calculation.  There is a high density of weakly
interacting gluons.

The initial energy density at the time $\tau \sim 1/\Lambda$
is
\be
	\epsilon \sim \Lambda^4 \sim A^{2/3}
\ee
At LHC energies, we will see that this corresponds to about
$100~ GeV/Fm^3$.

\section{Conclusions}

M. Gyulassy and I have recently estimated reasonable numbers
for the parameter $\chi$, the charge squared per unit
rapidity at rapidities greater than the center of mass rapidity
for hadron-hadron scattering.\cite{gyu}  This parameter sets the scale
of momentum.  We find that
\be
	\chi = \left( {A \over {200}} \right)^{1/3}
\left( {{10^{-2}} \over x} \right)^\delta (1 GeV)^2
\ee
where $\delta \sim 0.2$.  This follows from the Gluck-Reya-Vogt 
parameterization of structure functions.\cite{grv}

This corresponds to 300-400 MeV at RHIC energies and about 1 GeV at
LHC energies.  Neither of these numbers is large enough so that a 
a weak coupling treatment will be absolutely reliable.  It should
be semi-quantitative at LHC, and perhaps qualitative at RHIC.

If the ideas above can be tested semi-quantitatively and qualitatively,
then one will be confident one understands the physics.  Then at least at
asymptotically high energies, one will have a fundamental 
understanding of hadronic interactions.

\section{Acknowledgements}  

I thank my colleagues Alejandro Ayala-Mercado, Miklos Gyulassy,
Yuri Kovchegov, Alex Kovner, Jamal Jalilian-Marian, Andrei Leonidov,
Raju Venugopalan and Heribert Weigert with whom the ideas presented
in this talk were developed.  This work was supported under Department of
Energy grants in high energy and nuclear physics DOE-FG02-93ER-40764
and DOE-FG02-87-ER-40328.


\begin{thebibliography} {99}

\bibitem{mv}
L. McLerran and  R. Venugopalan, {\em Phys. Rev.} {\bf D49} 2233 (1994); 
{\bf D49} 3352 (1994).

\bibitem{glr}L. V. Gribov, E. M. Levin and M. G. Ryskin, {\it Phys. Rep.} 
{\bf 100} 1 (1983).


\bibitem{mk}J. Jalilian-Marian, A. Kovner, L. McLerran and H. Weigert, 
{\it Phys. Rev.} {\bf D55} 5414 (1997); J. Jalilian-Marian, A. Kovner,
A. Leonidov and H. Weigert hep-ph-9701284.


\bibitem{eqn}E.A. Kuraev, L.N. Lipatov and Y.S. Fadin, {\it Zh. Eksp. Teor. 
Fiz}
{\bf 72}, 3 (1977) ({\it Sov. Phys. JETP }{\bf 45}, 1 (1977) ); I.A.
Balitsky and L.N. Lipatov, {\it Sov. J. Nucl. Phys. }{\bf 28} 822 (1978); 
G. Altarelli and G. Parisi, {\it Nucl. Phys.} {\bf B126} 298 (1977); 
Yu.L. Dokshitser,  {\it Sov.Phys.JETP} {\bf 46} 641 (1977). 

\bibitem{mkw} {\it Phys. Rev. } {\bf D52}, 3809 (1995); 6231 (1995).

\bibitem{gyu} M. Gyulassy and L. McLerran, Columbia University Preprint, 
CU-TH-826.


\bibitem{grv}M. Gl\"uck, E. Reya and A. Vogt, {\it Z. Phys.} {\bf C67} 
433 (1995).


\end{thebibliography}
\end{document}